# Fractional and fractal derivatives modeling of turbulence

W. Chen

*Institute of Applied Physics and Computational Mathematics, P.O. Box 8009, Division Box 26, Beijing 100088, China*

This study makes the first attempt to use the 2/3-order fractional Laplacian modeling of enhanced diffusing movements of random turbulent particle resulting from nonlinear inertial interactions. A combined effect of the inertial interactions and the molecule Brownian diffusivities is found to be the bi-fractal mechanism behind multifractal scaling in the inertial range of scales of moderate Reynolds number turbulence. Accordingly, a stochastic equation is proposed to describe turbulence intermittency. The 2/3-order fractional Laplacian representation is also used to construct a fractional Reynolds equation for nonlinear interactions of fluctuating velocity components, underlying turbulence spacetime fractal structures of Lévy 2/3 stable distribution. The new perspective of this study is that the fractional calculus is an effective approach modeling of chaotic fractal phenomena induced by nonlinear interactions.

## 1. Introduction

The Kolmogorov -5/3 scaling characterizes the statistical similarity of turbulent motion at small scales based on the argument of local homogeneous isotropy [1]. To some extent, the scaling law has been validated by numerous experimental and numerical data of sufficiently high Reynolds number turbulence[1,2]. However, a clear departure from -5/3 scaling exponent is also often observed in various turbulence experiments at finite Reynolds numbers, namely, the so-called intermittency. The consensus is that the intermittent property of turbulence calls for a power law of energy spectrum having exponent $-5/3-c$ ($c \geq 0$). There are a few theories in the derivation of the correction exponent $c$. For instance, the $\beta$ model, various multifractal model[3,4], and Kolmogorov himself also refined his original -5/3 scaling by assuming that the kinetic energy dissipation rate $\varepsilon$ is scale-dependent and obeys a lognormal distribution leading to the so-called intermittency correction[1].

A school of researchers consider that the non-Gaussian distribution of turbulence velocity leads to the violation of the original Kolmogorov scaling and intermittency manifests in fact a non-Gaussian velocity distribution[4]. This argument has been controversial since many regard that the Kolmogorov theory does not assume the velocity Gaussianity. In section 2, we revisit this issue and show that the Kolmogorov scaling indeed underlies the Gaussian distribution of velocity increments while does not require Gaussianity of the displacement and acceleration fields. There exist quite a few statistical models of turbulence intermittency. To my knowledge, little, however, has been achieved in the partial differential equation modeling of intermittency. This study proposes a fractional Laplacian stochastic equation to describe the intermittency, whose one-dimension solution is given in appendix A. In section 3, by representing the nonlinear interactions of fluctuating velocity components with the fractional Laplacian, we obtain the fractional Reynolds equation underlying the Lévy 2/3 stable distribution of random turbulence displacements. Finally, section 4 concludes this paper with some remarks. In appendix B, the Richardson and Hausdorff fractal derivatives are used to model turbulence as alternative approaches to fractional derivatives. In appendix C, the elastic turbulence of non-Newtonian fluids is analyzed with the fractional time derivative. Appendix D proposes a revised cascade picture of turbulence energy transport.

The profound understanding of turbulence is up to now regarded as an unsolved problem. We consider that one major reason of this long-standing difficulty is the lacking of an appropriate mathematical devise. In this study, the innovative fractional calculus modeling is attempted to describe the complicated random phenomena of turbulence.

## 2. Intermittent statistical equation of turbulent diffusion

In the Kolmogorov's view of local homogeneous isotropic turbulence, the second-order structure function of velocity increments $\Delta u = u(x+r) - u(x)$ over a distance $r$ within the inertial range of scales is considered a stochastic variable and obeys a scaling law[5]

$$\left\langle (\Delta u)^2 \right\rangle = \hat{C}\varepsilon^{2/3} r^{2/3}, \quad \text{for } \eta \le r \le L_0, \tag{1}$$

where $\hat{C}$ is a universal dimensionless constant, the brackets represent the mean value of random variable ensemble, $\varepsilon$ denotes the kinetic energy dissipation rate per unit mass and is considered scale-independent, and $\eta = \left(\varepsilon^3/\upsilon\right)^{1/4}$ is the Kolmogorov dissipation length. The corresponding Kolmogorov scaling of turbulence kinetic energy transportation is

$$E(k) = C\varepsilon^{2/3} k^{-5/3}, \tag{2}$$

where $E(k)$ is the energy spectrum in terms of wavenumber $k$, and $C$ denotes the Kolmogorov constant. On the other hand, it is well known that the diffusion of displacements in the Kolmogorov turbulence is consistent with the Richardson's particle pair-distance superdiffusion[5-9] (enhanced diffusion) of a fully developed homogeneous turbulence, namely,

$$\langle r^2 \rangle = \overline{C} \varepsilon \Delta t^3, \qquad (3)$$

where $\Delta t$ denotes time interval and the experimental value of the dimensionless constant $\overline{C}$ is 0.5 given in ref. 6. (3) means particles move much faster than in normal diffusion ($\langle r^2 \rangle \propto \Delta t$). Through a dimensional analysis of (1) and (3), we can derive

$$\langle (\Delta u)^2 \rangle \propto \Delta t. \qquad (4)$$

The above (3) and (4) show that the Kolmogorov turbulence in the inertial range of scales is of the normal diffusion of the velocity difference and the enhanced diffusion of displacements. Consequently, turbulence in the inertial range is considered to have Gaussian velocity field and non-Gaussian displacement field. Laboratory experiments and field observations have found that the statistics of the velocity increments in the inertial range is often close to Gaussian[6]. The displacement diffusion (4) can be restated as

$$\langle r^2 \rangle \propto \Delta t^{2/\alpha}, \qquad \alpha = 2/3. \qquad (5)$$

The above formula (5) can be interpreted the displacement increments in turbulence obeys the Lévy $\alpha$-stable distribution[11], where $\alpha$ represents the stability index of the Lévy distribution. The rigorous mathematics proof shows that Lévy stability index $\alpha$ must be positive and not larger than 2 ($0<\alpha\leq 2$) with the Gaussian distribution being its limiting $\alpha =2$ case[10,11]. The non-Gaussian Lévy stable distribution of velocity difference has an algebraic decay tail. It is noted that the Gaussian distribution drastically underestimates the occurrence probability of the large events, while for heavy tailed statistics like the Lévy stable distribution the occurrence of extreme events is drastically enhanced.

The Lévy distribution has long been used to describe strong long-range spatiotemporal correlation, featuring heavy tails, of anomalous diffusion in turbulence[12-14]. To the author's knowledge, the corresponding differential equation model, however, has been missing. The fractional Laplacian has been a popular approach in recent years to model the Lévy statistical superdiffusion in a variety of physical master equations such as the Fokker-Planck equation[15] and the anomalous diffusion equation[10,16]. Intuitively, we construct a linear phenomenological statistical equation within the inertial range of scales of fully developed isotropic homogeneous turbulence at sufficiently high Reynolds numbers

$$\frac{\partial P}{\partial t} + \gamma(-\Delta)^{\alpha/2} P = 0, \qquad \alpha = 2/3, \qquad (6)$$

where $P(x,t)$ is the probability density function (pdf) to find a particle at $x$ at time instant $t$, and $(-\Delta)^{\alpha/2}$ represents the homogeneous symmetric (isotropic) fractional Laplacian[17,18], and $\gamma$ is turbulent diffusion coefficient. $\alpha$ can be understood the fractal dimension in this study. In terms of the generalized Einstein dissipation-fluctuation theorem[19] and (3), we can derive $\gamma = (\overline{C}\varepsilon/2)^{1/3}$. The Green function of the Cauchy problem of equation (6) results in the time-dependent Lévy pdf, which can naturally leads to Richardson's turbulence superdiffusion (3) through the so-called Lévy walk mechanism[7,12], while underlying the Gaussian velocity field and the Kolmogorov scaling in the inertial range of scales.

It is known that the scalings (3) and (4) are obtained for fully developed homogeneous turbulence under sufficiently high Reynolds numbers and reflect the statistical self-similarity of eddy structures generated from nonlinear inertial interactions. Therefore, the superdiffusion diffusion equation (6) actually describes the enhanced diffusivity originating from coarse-grained average of the nonlinear inertial term in the Navier-Stokes equation. And there is no advective term in (6). It is also noted that (6) is a linear phenomenological model equation to characterize the fractal self-similarity of complicated nonlinear interactions. We call equation (6) inertial diffusion equation.

On the other hand, for the finite Reynolds number turbulence, a clear deviation from Gaussian velocity field and $t^3$ displacement superdiffusion at small scales has been observed in various turbulence experiments and numerical simulations, namely, turbulence intermittency[20,21]. In the absence of molecular diffusion, model equation (6) can not describe the intermittency. The addition of molecular diffusion will reflect intermittency for finite Reynolds number turbulence, namely,

$$\frac{\partial P}{\partial t} + \gamma(-\Delta)^{1/3} P - \upsilon\Delta P = 0, \qquad (7)$$

where $\Delta$ represents the Laplacian operator, and $\upsilon$ molecular viscosity. (7) is called intermittent stochastic equation in this study. It is noted that the two diffusion terms in (7) are induced by the inertial interactions and molecular viscosity in the Navier-Stokes equation of motion, respectively, reflecting the two inherent physical systems behind stochastic turbulence phenomena. A space Fourier transform of (7) results in

$$\frac{\partial P(k,t)}{\partial t} + \gamma k^{2/3} P(k,t) + \upsilon k^2 P(k,t) = 0. \tag{8}$$

Then we have the probability characteristic function

$$P(k,t) = \exp(-\gamma |k|^{2/3} - \upsilon |k|^2)t. \tag{9}$$

pdf $P(x,t)$ can be evaluated by an inverse Fourier transform. It is apparent that the appearance of the molecular diffusivity in (6) destroys the Richardson displacement diffusions. Appendix A gives the solution of the one-dimensional equation (7).

Compared with Kraichnan's direct-interaction approximation (DIA) theory[22], (7) describes a phenomenological linear stochastic turbulence field in the presence of molecular diffusivity, while the DIA considers turbulence a nonlinear stochastic field. The Green functions of these two approaches are the statistical distribution of turbulence. However, the DIA is mathematically very complicated thanks to its nonlinearity, while the present fractional Laplacian model captures major fractal feature of nonlinear inertial interactions with merits being mathematically far simpler and easier to evaluate.

To measure the intermittency, we introduce the dimensional ratio value

$$\theta = \gamma / \upsilon k^{4/3}. \tag{10}$$

The inertial diffusivity $\gamma$ is considered much larger compared with the molecular diffusivity $\upsilon$. However, we note from (10) that the extent of intermittency is also dependent on wavenumber and increases with it. Through a dimensional analysis, we find $\theta \propto \text{Re}^{2/3}/(Lk)^{4/3}$, where $Re$ and $L$ are Reynolds number and characteristic length of fluid flows, respectively. The larger the Reynolds number, the more dominant is the inertial diffusion. This conforms the consensus that the intermittency is of Reynolds-number-dependency.

For fully developed isotropic homogeneous turbulence at sufficiently high Reynolds numbers, the intermittency parameter $\theta$ is very large for relatively low wavenumbers in the so-called convective-inertial range. And the molecular diffusivity vanishes or is small enough that its effects are negligible. Consequently, equation (7) is reduced to the limiting equation (6) and the turbulence displacement is dominated by the third power of time law (3) and the Gaussian velocity field in the classical K41 inertial range. On another extreme limit when the value of $\theta$ is very small for very high wavenumbers towards molecular scales, the inertial diffusivity is relatively weak to be neglected and equation (7) is reduced to the normal diffusion equation for molecular Brownian motion. And in this case the turbulence displacement field is Gaussian

$$\langle r^2 \rangle = 2\upsilon t. \tag{11}$$

while the velocity field is described by[4]

$$\langle (\Delta u)^2 \rangle = A\frac{\varepsilon}{\upsilon}r^2. \tag{12}$$

It is clear that the displacement fields vary from the 2/3 Lévy stable distribution to the Gaussian distribution. The nonlinear inertial interactions yield the Gaussian velocity field and the 2/3 Lévy stable distribution displacement, while the molecular viscosity is responsible for the non-Gaussian velocity field and the Gaussian displacement. The resulting velocity and displacement fields are a combined effect of these two contributing sources[23] to display varied intermittency deviations. Therefore, the inertial range is split into the two parts: 1) the convective-inertial range where the inertial interaction diffusion dominates and intermittency is less apparent; 2) the inertial-viscous range where the molecular diffusion can not be neglected and intermittency is observed.

The power spectral of kinetic energy of turbulence obeys the scaling law

$$E(k) \propto k^{-\beta}, \tag{13}$$

where the exponent parameter $\beta$ has a simple relation with the exponents $q$ of the corresponding second moment ($\langle (\Delta u)^2 \rangle \propto r^q$) of velocity random fields[24]: $\beta=q+1$. For Kolmogorov Gaussian velocity field (1), $q=2/3$ and $\beta=5/3$, while for velocity field (12), $q=2$ and $\beta=3$. The value of $\beta$ between these two extreme cases ranges from 5/3 to 3 as $q$ varies from 2/3 to 2 as a function of wavenumber. For instance, $q=1$ and $\beta=2$. Thus, turbulence scaling is of multifractal in nature. The stochastic essence of turbulence flows is its diffusion behaviors.

**3. Reynolds equation model with fractional derivative**

The intermittent statistical equation (7) can be considered the fractional Fokker-Planck equation (FFPE) with the fractional Laplacian to describe anomalous diffusion. It is well known that the classical Fokker-Planck equation underlies the classical Navier-Stokes equation. This inspires us to apply the fractional representation in the Navier-Stokes equation modeling of turbulence. The equations of motion of an incompressible fluid are

$$\frac{\partial \mathbf{u}}{\partial t} + \mathbf{u} \cdot \nabla \mathbf{u} = -\frac{1}{\rho}\nabla p + \upsilon \Delta \mathbf{u}, \tag{14a}$$

$$\nabla \cdot \mathbf{u} = 0, \tag{14b}$$

where **u** is the velocity vector and *p* represents pressure. Following Reynolds, velocity and pressure can be decomposed as a sum of mean flow components $\bar{u}$, $\bar{p}$ and small-scale fluctuating components $\tilde{u}$, $\tilde{p}$. The mean value of fluctuating quantities are considered zero. Substituting the decomposition of velocity and pressure into equations (14), we have the following Reynolds equations[25]

$$\frac{\partial \bar{u}_i}{\partial t} + \bar{u}_j \cdot \frac{\partial \bar{u}_i}{\partial x_j} = -\frac{1}{\rho}\nabla \bar{p} + \upsilon \Delta \bar{u}_i - \frac{\partial}{\partial x_j}\langle \tilde{u}_i \tilde{u}_j \rangle, \tag{15a}$$

$$\nabla \cdot \bar{u}_i = 0. \tag{15b}$$

The nonlinear fluctuation term $\partial \langle u_i u_j \rangle / \partial x_j$ gives rise to the controversial closure problem in the Reynolds equations. For the fully developed homogeneous isotropic turbulence, the fluctuating velocity components are considered to exhibit a variety of universal features of statistically homogeneous isotropy and self-similar eddy structures, corresponding to the Richardson and Kolmogorov's picture of cascade transport of kinetic energy in inertial range of scales. Intermittency is interpreted as the joint action of the mean zero random velocity field and molecular diffusion on the large scale and long times. By analogy with the previous statistical equation (7), we present a representation of these universal characteristics of the Reynolds nonlinear fluctuation interactions by

$$\frac{\partial}{\partial x_j}\langle \tilde{u}_i \tilde{u}_j \rangle = \gamma(-\Delta)^{1/3} \bar{u}_i. \tag{16}$$

(16) can be considered the turbulence diffusivity which leads to the enhanced diffusion. It is noted that (16) is different from the so-called eddy (effective) diffusivity of empirical turbulence models in that it underlies Gaussian velocity and the Kolmogorov scaling. The turbulence viscosity (16) is in agreement with the Kolmogorov's key hypothesis that the small-scale structures of turbulence flows, away from boundaries, are largely independent of the large scale configuration. Then we have the fractional derivative Reynolds equation

$$\frac{\partial \bar{u}_i}{\partial t} + \bar{u}_j \cdot \frac{\partial \bar{u}_i}{\partial x_j} = -\frac{1}{\rho}\nabla \bar{p} + \upsilon \Delta \bar{u}_i - \gamma(-\Delta)^{1/3} \bar{u}_i. \tag{17}$$

Here the fractional Laplacian $(-\Delta)^{1/3}$ serves as a stochastic driver underlying statistical self-similarity in the inertial range and guarantees the positive definiteness of energy dissipation. The molecular diffusivity is a property of fluids, while the inertial diffusivity is a characteristic of flows[25] which reflect the long-range correlation (memory) in turbulence chaotic motions, apparently resembling an inherent property of non-Newtonian fluids. In other words, the fractional Laplacian representation is to describe the complicated flow property rather than complex fluid constitutive relationship. In appendix D, we will discuss the similarity in phenomenological descriptions of complex fluids and turbulence motions.

The renormalization group technique may be a plausible approach to derive (17) directly from the Navier-Stokes equation. It is worth mentioning that the naive numerical solution of the fractional Reynolds equation will be computationally expensive, since the fractional Laplacian are a non-local operator[17,18] and will result in the full matrix of numerical discretization[26]. The fast algorithms based on the preconditioning techniques such as the fast multipole method, panel clustering, and H-matrix method will be of vital importance to perform effective numerical simulations.

Let $T$, $L$, $V_\infty$, and $P$ represent the characteristic time, length, velocity, and pressure of the fluid flow, we can then have the dimensionless expression of the fractional Reynolds equation (17)

$$St \frac{\partial \overline{\mathbf{u}}^*}{\partial t} + \overline{\mathbf{u}}^* \cdot \nabla \overline{\mathbf{u}}^* = -Et \nabla \overline{p}^* + \frac{1}{Re} \Delta \overline{\mathbf{u}}^* - (0.5\overline{C})^{1/3} \frac{1}{Re^{1/3}} (-\Delta)^{1/3} \overline{\mathbf{u}}^*. \quad (18)$$

where $\overline{\mathbf{u}}^*$ and $\overline{p}^*$ are dimensionless velocity and pressure. $St$ is a constant, $Et$ denotes the Euler number, and $Re$ represents the Reynolds number. $(0.5\overline{C})^{1/3} \approx 0.63$. (18) shows that the coefficient of the inertial chaos diffusion is three orders of magnitude greater than that of molecular diffusion. For instance, the inertial diffusion constant is only 100 in a Reynolds number $10^6$ flow.

The equation of motions (17) is deterministic, but their solutions have many attributes of random processes thanks to both of the Laplacian and fractional Laplacian viscous terms. We also find (17) satisfies the same scale invariance of the standard Navier-Stokes equation[1,4]

$$x' = \lambda x, \quad t' = \lambda^{2/3} t, \quad u' = \lambda^{1/3} u,$$

$$\left(p/\rho\right)' = \lambda^{2/3}\left(p/\rho\right), \quad \upsilon' = \lambda^{4/3}\upsilon, \quad \gamma' = \gamma. \tag{19}$$

The very nature of fractional Laplacian representation in the present Reynolds equations (18) also underlies the ballistic motion of turbulence particles under the Lévy walk picture[7] and implies that the turbulence diffusion is not fully irreversible, in between deterministic advection and fully random (irreversible) diffusion motions[27] and may have stochastic and deterministic duality versus the convectional fully irreversible stochastic turbulence. In essence, this study presents a simple mathematical formulation of chaos in which the deterministic Newtonian dynamics generates the random thermoviscous behavior.

**4. Concluding remarks**

By using the fractional Laplacian, this paper presents new statistical-mechanical descriptions of the dynamics of chaos-induced turbulence diffusion. The standard chaos dynamics models are mainly characterized by temporal complexity, but spatial simplicity[4], while the present fractional equations can be considered a kind of fractal continuum dynamics, complex both in time and space. The fractional calculus, fractal, and Lévy distribution are consistent mathematical concepts to describe complicated dissipation, transport, and diffusion phenomena of turbulence. One of major new perspective in this study is that the fractional calculus is an effective approach to model the fractal phenomena resulting from chaotic nonlinear interactions. Although the nonlinear partial differential description and the fractional derivative representation are seemingly quite different mathematical approaches and the underlying relationship between them is still not explicit, their common feature is fractal in statistical physics which leads to fractional calculus modeling of chaos induced diffusions.

Turbulence intermittency has long been considered to possess multifractal structures[3]. However, to my best knowledge, an explicit partial differential equation of multifractal does not exist and the multifractal mechanism is not well established. As discussed in section 2, the present dual diffusivity model equations provide a clear picture of how the displacement field distributions vary with wavenumber and fluid molecule viscosity. The bi-fractal model of 2/3 fractional Laplacian inertial diffusion and molecule viscosity generates multifractal in turbulence.

Warhaft[28] pointed out "Apart from noting the presence of non-Gaussian tails, no deeper analysis of the shape of the *pdf*s has been made. Because the connection of these models to the Navier-Stokes equations is tenuous,…". In this study, the attempt was also made to explicitly connect non-Gaussian statistics of turbulence and the Reynolds equation, a variant of the Navier-Stokes equation, where the fractional Laplacian representation describes the kinetic energy transportation and dissipation induced by the complex nonlinear interactions. Section 3 actually presents a statistical and physics model of the closure of the Reynolds equation.

It is worth mentioning that the fractional Laplacian[17,18] and Lévy stable distribution[11] can be asymmetric to describe the skewness of turbulence distributions and ballistic motion[7], which this study does not touch on. On the other hand, the stretched Gaussian[29,30] and Hausdorff derivatives[30] can also properly describe anomalous diffusion. Thus, the Lévy stable distribution and fractional derivatives may not be the only approaches in modeling turbulence. Appendix B presents the alternative models via the fractal Hausdorff and Richardson derivatives.

**Acknowledgments**:

The author is grateful of helpful discussions with Dr. H.B. Zhou in the preparation of this paper.

**Appendix A: The solution of one-dimensional intermittent stochastic equation**

We consider one-dimensional Cauchy problem of intermittent stochastic equation (7)

$$\frac{\partial P}{\partial t} - \gamma \frac{\partial^{2/3} P}{\partial |x|^{2/3}} - \upsilon \frac{\partial^2 P}{\partial x^2} = 0, \quad \text{(A1)}$$

$$P(x,0) = P_0(x). \quad \text{(A2)}$$

Rewriting the above equation (A1) with dimensionless $\hat{x} = x/\eta$, $\hat{t} = t/\tau$, where $\eta = (\upsilon^3/\varepsilon)^{1/4}$ and $\tau = (\upsilon/\varepsilon)^{1/2}$ are the Kolmogorov dissipation length and time, respectively, we have

$$\frac{\partial P}{\partial \hat{t}} - \mu \frac{\partial^{2/3} P}{\partial |\hat{x}|^{2/3}} - \frac{\partial^2 P}{\partial \hat{x}^2} = 0, \quad \text{(A3)}$$

$$P(\hat{x},0) = P_0(\hat{x}). \quad \text{(A4)}$$

where $\mu = (\overline{C}/2)^{1/3}$ is a dimensionless constant. By using the Fourier transform[31], we has the solution

$$P(\hat{x},\hat{t}) = p_{2/3}(\hat{t}) * p_2(\hat{t}) * P_0(\hat{x}), \tag{A5}$$

where

$$p_{2/3}(\hat{x},\hat{t}) = (2\pi)^{-1} \int_{-\infty}^{+\infty} e^{-\hat{t}\mu|\hat{k}|^{2/3} + i\hat{x}\hat{k}} d\hat{k}$$

and

$$p_2(\hat{x},\hat{t}) = (4\pi\hat{t})^{-1/2} \exp\left(-|\hat{x}|^2/(4\hat{t})\right).$$

**Appendix B: Hausdorff and Richardson fractal derivative models**

Recently, the present author introduces the Hausdorff derivative for modeling of fractal underlying problems[30], which is defined by

$$\Delta_F^{\alpha/2} P = \frac{\partial}{\partial x^{\alpha/2}} \frac{\partial P}{\partial x^{\alpha/2}}. \tag{B1}$$

The Hausdorff derivative can be used to build the partial differential equation model of anomalous diffusion, underlying the stretched Gaussian distribution and the fractional Brownian motion, widely used in fitting turbulence data. In comparison to the previous intermittent stochastic equation (7) of fractional derivative, we can have the corresponding Hausdorff derivative expression

$$\frac{\partial P}{\partial t} - \kappa_0 \Delta_F^{1/3} P - \upsilon \Delta P = 0, \tag{B2}$$

where $\kappa_0$ denotes the diffusivity constant. Obviously, $\Delta_F^1 P = \Delta P$. By analogy with the fractional derivative Reynolds equation (17), we also have a new expression of fractional Reynolds equation as follows

$$\frac{\partial \overline{u}_i}{\partial t} + \overline{u}_j \cdot \frac{\partial \overline{u}_i}{\partial x_j} = -\frac{1}{\rho}\nabla \overline{p} + \upsilon \Delta \overline{u}_i + \kappa_0 \Delta_F^{1/3} \overline{u}_i. \tag{B3}$$

On the other hand, in order to model the displacement superdiffusion, Richardson constructed an empirical phase-space statistical equation of fully developed homogeneous turbulence, i.e., the so-called Richardson 4/3 law[1,5-7],

$$\frac{\partial P(r,t)}{\partial t} - \frac{\partial}{\partial r}\left(\kappa(r)\frac{\partial P(r,t)}{\partial r}\right) = 0, \tag{B4}$$

where the turbulent diffusion $\kappa(r) \propto r^{4/3}$, and $P(r,t)$ is the probability that the two particles, initially close together, have a relative separation $r$ at time $t$. The Green function of (B4) is an extended Gaussian distribution[5-7], and leads to the displacement superdiffusion (3). (B4) is a model competitive to inertial diffusion equation (6). Likewise, we can extend the Richardson model to the inclusion of the molecule diffusivity, namely,

$$\frac{\partial P}{\partial t} - \kappa_0 \Delta_R^{1/3} P - \upsilon \Delta P = 0, \tag{B5}$$

where we introduce $\Delta_R^{\alpha/2} P = \frac{\partial}{\partial x}\left(x^{(2-\alpha)}\frac{\partial P}{\partial x}\right)$ as a new type of fractal derivative and called Richardson fractal derivative, $\Delta_R^1 P = \Delta P$. By analogy with the fractional derivative Reynolds equation (17), we can also have a new expression of fractional Reynolds equation.

The advantage of the present Hausdorff and Richardson equation models is easier to compute than the previous fractional derivative model equations thanks to their local property. In addition, the stretched Gaussian is much more frequently used in turbulence modeling than the Lévy distribution. The links and differences between these different differential expressions and the underlying statistical distributions are still an open issue[7].

**Appendix C: Fractional Brownian motion modeling of elastic turbulence of non-Newtonian fluids**

The definition of anomalous diffusion is based only on the time evolution of the mean square displacement of diffusing turbulent particle movements[11, 17,18]

$$\langle r^2 \rangle \propto \Delta t^\sigma. \tag{C1}$$

For anomalous diffusion ($\sigma \neq 1$), particles move coherently for long times with infrequent changes of direction, faster in superdiffusion ($\sigma \succ 1$, e.g., $\sigma=3$ in (3) for enhanced turbulence diffusion in the inertial range) and more slowly in subdiffusion ($\sigma \prec 1$) than linearly with time in normal diffusion ($\sigma = 1$), i.e., $\langle r^2 \rangle \propto \Delta t$. Both superdiffusion and subdiffusion turbulences have been observed. This paper has by now focused on the superdiffusion in the inertial range. This appendix is concerned with modeling of subdiffusion particle movements underlying statistics of the fractional Brownian motion in non-Newtonian fluids, whose constitutive relationship can be modeled by the fractional time derivative

$$s = \hat{\upsilon}\frac{\partial^\lambda}{\partial t^\lambda}\nabla u, \quad 0 \leq \lambda < 1, \tag{C2}$$

where $\hat{\upsilon}$ is the diffusivity constant of non-Newtonian fluid macromolecules, and $\lambda$ is a material parameter and the memory strength index, and the larger $\lambda$, the stronger memory. The above constitutive equation (C2) reflects the history-dependent motion, e.g., the elastic turbulence of polymer solutions[32]. Compared with (7), the corresponding intermittent stochastic equation is given by

$$\frac{\partial P}{\partial t} + \gamma(-\Delta)^{1/3} P - \hat{\upsilon}\frac{\partial^\lambda}{\partial t^\lambda}\Delta P = 0. \tag{C3}$$

Replacing the standard molecular viscosity in the fractional Laplacian Reynold equation (17), we have

$$\frac{\partial \bar{u}_i}{\partial t} + \bar{u}_j \cdot \frac{\partial \bar{u}_i}{\partial x_j} = -\frac{1}{\rho}\nabla \bar{p} + \hat{\upsilon}\frac{\partial^\lambda}{\partial t^\lambda}\Delta \bar{u}_i - \gamma(-\Delta)^{1/3}\bar{u}_i. \tag{C4}$$

It is noted that subdiffusion turbulence particles displacements are not a Markovian process and can be explained by the so-called continuous time random walk theory[33].

In addition, the fractional time derivative can also be used to model the fractal diffusivity of nonlinear inertial interactions, namely, we can use $\hat{\gamma}\partial^{\frac{2}{3}}\bar{u}_i \big/ \partial t^{\frac{2}{3}}$ to replace $\gamma(-\Delta)^{1/3}\bar{u}_i$.

**Appendix D: Conjectures on energy dissipation and velocity distributions**

In recent decades complex fluids, also known as soft matters such as oil, biomaterials, sediments, polymer solutions, etc., has attracted a lot of attention across a host of research areas[34]. Thermoviscous dissipation of acoustic wave propagation through complex fluids is described by

$$E = E_0 e^{-\xi(\omega)z}, \tag{D1}$$

where $E$ represents energy, $z$ is the traveling distance, and $\omega$ denotes angular frequency. The dissipation coefficient $\xi$ characterizes an empirical power law function over a finite but broad range of frequency[35-37]

$$\xi = \xi_0 |\omega|^y, \qquad 0 \prec y \leq 2 \qquad (D2)$$

where $\xi_0$ and $y$ are non-negative media-dependent constants. In essence, exponent $y$ describes the fractal nature of complex fluids and underlies their mechanics constitutive relationship. It is noted that physical properties of complex fluids can not arise from relativistic or quantum properties of elementary molecules and are determined by many-body interactions such as entanglements, branching, breaking, and cross-linking (e.g., polymer macromolecules). The fractional calculus and the Lévy statistics have been successfully used to describe and model physical behaviors of complex fluids.

In turbulence, the large amount of the elementary molecules is grouped together to form eddy structures on scales generally much larger than a single molecule, which entangle each other and grow, collide and break into small vortices, diffuse and dissipate energy, and behave like dissipative macromolecules in complex fluids. The nonlinear inertial interactions dominates turbulence transport property and fractal eddy structures in the inertial range of scales which has been characterized by the 1/3 fractional Laplacian (or Richardson and Hausdorff fractal derivative) representation in this study. As observed in ref. 12, the behaviors of complicated fluid flows in turbulence appear like those of complex fluids, despite the fact that turbulence eddies are much less stable than macromolecules of complex fluids. The mechanism behind turbulence scaling laws is considered comparable to the frequency power dissipation (D2) of complex fluids, both of which have found to obey the Lévy stable process or the stretched Gaussian distribution.

Based on the above arguments, we conjecture that (D1) also holds effective for dissipative motion of a turbulence particle and shows that the energy is dissipated across a finite range of scale (frequency). This is in sharp contrast to the fundamental assumption in the Richardson and Kolmogorov's cascade picture that large eddies transmit their energy without dissipation to smaller eddies down to the Kolmogorov scale, only where energy is dissipated by molecule viscosity. According to (D1), a revised cascade picture of turbulence appears that the energy is dissipated by eddy viscosity through all its downward scale transport. This means that the so-called inertial range in turbulence is of convection-dissipation process rather than a pure convection (transport) process. A new physical interpretation of turbulence will be a contentious issue. The presented mechanism of turbulence transport and dissipation is of speculative nature. This mechanism can be examined by studying how the turbulence responds to a narrow band of forcing frequencies.

It is also noted that the scaling exponent *y* in (D2) for different complex fluids ranges from 0 to 2 corresponding different Lévy stable processes with varying stability index. By comparison, we conjecture that the turbulence velocity fields endure all Lévy stable distribution from Gaussian to Lévy stable distribution with almost zero stability index which appears like the log-normal distribution in its central part, remaindering of Kolmogorov's 1962 modifications.